\documentclass{aastex631}
\usepackage{amsmath}
\usepackage{threeparttable}

\begin{document}

\title{Sudden polarization angle jumps of the repeating fast radio burst FRB 20201124A}

\author[0000-0001-8065-4191]{J.~R.~Niu}
\affiliation{National Astronomical Observatories, Chinese Academy of Sciences, 20A Datun Road, Chaoyang District, Beijing 100101, China}
\affiliation{School of Astronomy and Space Science, University of Chinese Academy of Sciences, Beijing 100049, China} 

\author[0000-0001-9036-8543]{W.~Y.~Wang} 
\affiliation{School of Astronomy and Space Science, University of Chinese Academy of Sciences, Beijing 100049, China} 

\author{J.~C.~Jiang}
\affiliation{National Astronomical Observatories, Chinese Academy of Sciences, 20A Datun Road, Chaoyang District, Beijing 100101, China}

\author[0000-0003-4721-4869]{Y.~Qu}
\affiliation{The Nevada Center for Astrophysics, University of Nevada, Las Vegas, NV 89154, USA}
\affiliation{Department of Physics and Astronomy, University of Nevada, Las Vegas, NV 89154, USA}

\author[0000-0002-6423-6106]{D.~J.~Zhou}
\affiliation{National Astronomical Observatories, Chinese Academy of Sciences, 20A Datun Road, Chaoyang District, Beijing 100101, China}

\author[0000-0001-5105-4058]{W.~W.~Zhu} 
\affiliation{National Astronomical Observatories, Chinese Academy of Sciences, 20A Datun Road, Chaoyang District, Beijing 100101, China}
\email{zhuww@nao.cas.cn}

\author{K.~J.~Lee} 
\affiliation{National Astronomical Observatories, Chinese Academy of Sciences, 20A Datun Road, Chaoyang District, Beijing 100101, China}
\affiliation{Kavli Institute for Astronomy and Astrophysics, Peking University, Beijing 100871, China}
\affiliation{Department of Astronomy, Peking University, Beijing 100871, China}
\email{kjlee@pku.edu.cn}

\author[0000-0002-9274-3092]{J.~L.~Han} 
\affiliation{National Astronomical Observatories, Chinese Academy of Sciences, 20A Datun Road, Chaoyang District, Beijing 100101, China}
\email{hjl@nao.cas.cn}

\author[0000-0002-9725-2524]{B.~Zhang}
\affiliation{The Nevada Center for Astrophysics, University of Nevada, Las Vegas, NV 89154, USA}
\affiliation{Department of Physics and Astronomy, University of Nevada, Las Vegas, NV 89154, USA}
\email{bing.zhang@unlv.edu}

\author{D.~Li}
\affiliation{National Astronomical Observatories, Chinese Academy of Sciences, 20A Datun Road, Chaoyang District, Beijing 100101, China}

\author{S.~Cao}
\affiliation{Yunnan Observatories, Chinese Academy of Sciences, P.O.Box110, Kunming 650011, Yunnan, China} 
\affiliation{School of Astronomy and Space Science, University of Chinese Academy of Sciences, Beijing 100049, China} 

\author{Z.~Y.~Fang}
\affiliation{National Astronomical Observatories, Chinese Academy of Sciences, 20A Datun Road, Chaoyang District, Beijing 100101, China}

\author{Y.~Feng}
\affiliation{Zhejiang Lab, Kechuang Avenue, Zhongtai Sub-District, Yuhang District, Hangzhou, Zhejiang Province,  China}

\author{Q.~Y.~Fu}
\affiliation{National Astronomical Observatories, Chinese Academy of Sciences, 20A Datun Road, Chaoyang District, Beijing 100101, China}
\affiliation{School of Astronomy and Space Science, University of Chinese Academy of Sciences, Beijing 100049, China} 

\author{P.~Jiang}
\affiliation{National Astronomical Observatories, Chinese Academy of Sciences, 20A Datun Road, Chaoyang District, Beijing 100101, China}

\author{W.~C.~Jing}
\affiliation{National Astronomical Observatories, Chinese Academy of Sciences, 20A Datun Road, Chaoyang District, Beijing 100101, China}
\affiliation{School of Astronomy and Space Science, University of Chinese Academy of Sciences, Beijing 100049, China} 

\author{J.~Li}
\affiliation{School of Computer Science and Engineering, Beihang University, Beijing, 100191, China}

\author{Y.~Li}
\affiliation{Purple Mountain Observatory, Chinese Academy of Sciences, Nanjing 210023, China}

\author{R.~Luo}
\affiliation{Department of Astronomy, School of Physics and Materials Science, Guangzhou University, Guangzhou 510006, China}

\author{L.~Q.~Meng}
\affiliation{National Astronomical Observatories, Chinese Academy of Sciences, 20A Datun Road, Chaoyang District, Beijing 100101, China}
\affiliation{School of Astronomy and Space Science, University of Chinese Academy of Sciences, Beijing 100049, China} 

\author{C.~C.~Miao}
\affiliation{Zhejiang Lab, Kechuang Avenue, Zhongtai Sub-District, Yuhang District, Hangzhou, Zhejiang Province,  China}
 
\author{X.~L.~Miao}
\affiliation{National Astronomical Observatories, Chinese Academy of Sciences, 20A Datun Road, Chaoyang District, Beijing 100101, China}

\author{C.~H.~Niu}
\affiliation{Institute of Astrophysics, Central China Normal University, Wuhan 430079, China }

\author{Y.~C.~Pan}
\affiliation{Department of Astronomy, Peking University, Beijing 100871, China}

\author{B.~J.~Wang}
\affiliation{National Astronomical Observatories, Chinese Academy of Sciences, 20A Datun Road, Chaoyang District, Beijing 100101, China}

\author{F.~Y.~Wang}
\affiliation{School of Astronomy and Space Science, Nanjing University, Nanjing 210093, China}
\affiliation{Key Laboratory of Modern Astronomy and Astrophysics (Nanjing University), Ministry of Education, China}

\author{H.~Z.~Wang}
\affiliation{Lanzhou University, Lanzhou 730000, China}
 
\author{P.~Wang}
\affiliation{National Astronomical Observatories, Chinese Academy of Sciences, 20A Datun Road, Chaoyang District, Beijing 100101, China}

\author{Q.~Wu}
\affiliation{School of Astronomy and Space Science, Nanjing University, Nanjing 210093, China}
\affiliation{Key Laboratory of Modern Astronomy and Astrophysics (Nanjing University), Ministry of Education, China}

\author{Z.~W.~Wu}
\affiliation{National Astronomical Observatories, Chinese Academy of Sciences, 20A Datun Road, Chaoyang District, Beijing 100101, China}

\author{H.~Xu}
\affiliation{National Astronomical Observatories, Chinese Academy of Sciences, 20A Datun Road, Chaoyang District, Beijing 100101, China}

\author{J.~W.~Xu}
\affiliation{Department of Astronomy, Peking University, Beijing 100871, China} 
\affiliation{Kavli Institute for Astronomy and Astrophysics, Peking University, Beijing 100871, China}
 
\author{L.~Xu}
\affiliation{Faculty of Electrical engineering and Computer Science, Ningbo University, Ningbo 315211, China}
\affiliation{Pengcheng Laboratory, Shenzhen, 518055, China}
  
\author{M.~Y.~Xue}
\affiliation{National Astronomical Observatories, Chinese Academy of Sciences, 20A Datun Road, Chaoyang District, Beijing 100101, China}

\author[0000-0001-6374-8313]{Y.~P.~Yang} 
\affiliation{South-Western Institute for Astronomy Research, Yunnan University, Kunming, Yunnan 650504, China}
\affiliation{Purple Mountain Observatory, Chinese Academy of Sciences, Nanjing 210023, China} 

\author{M.~Yuan}
\affiliation{National Space Science Center, Chinese Academy of Sciences, Beijing 100190, China}
 
\author{Y.~L.~Yue}
\affiliation{National Astronomical Observatories, Chinese Academy of Sciences, 20A Datun Road, Chaoyang District, Beijing 100101, China}

\author{D.~Zhao}
\affiliation{National Space Science Center, Chinese Academy of Sciences, Beijing 100190, China}

\author{C.~F.~Zhang}
\affiliation{National Astronomical Observatories, Chinese Academy of Sciences, 20A Datun Road, Chaoyang District, Beijing 100101, China}

\author{D.~D.~Zhang}
\affiliation{School of Physics and Electronic Science, Guizhou Normal University, Guiyang 550025, China}
\affiliation{Guizhou Provincial Key Laboratory of Radio Astronomy and Data Processing, Guizhou Normal University, Guiyang 550025, China}

\author{J.~S.~Zhang}
\affiliation{National Astronomical Observatories, Chinese Academy of Sciences, 20A Datun Road, Chaoyang District, Beijing 100101, China}
\affiliation{School of Astronomy and Space Science, University of Chinese Academy of Sciences, Beijing 100049, China} 

\author{S.~B.~Zhang}
\affiliation{Purple Mountain Observatory, Chinese Academy of Sciences, Nanjing 210023, China} 
 
\author[0000-0002-8744-3546]{Y.~K.~Zhang}
\affiliation{National Astronomical Observatories, Chinese Academy of Sciences, 20A Datun Road, Chaoyang District, Beijing 100101, China}
\affiliation{School of Astronomy and Space Science, University of Chinese Academy of Sciences, Beijing 100049, China}

\author{Y.~H.~Zhu}
\affiliation{National Astronomical Observatories, Chinese Academy of Sciences, 20A Datun Road, Chaoyang District, Beijing 100101, China}
\affiliation{School of Astronomy and Space Science, University of Chinese Academy of Sciences, Beijing 100049, China}








\begin{abstract}
We report the first detection of polarization angle (PA) orthogonal jumps, a phenomenon previously only observed from radio pulsars, from a fast radio burst (FRB) source FRB 20201124A.
We find three cases of orthogonal jumps in over two thousand bursts, all resembling those observed in pulsar single pulses. We propose that the jumps are due to the superposition of two orthogonal emission modes that could only be produced in a highly magnetized plasma, and they are caused by the line of sight sweeping across a rotating magnetosphere. 
The shortest jump timescale is of the order of one-millisecond, which hints that the emission modes come from regions smaller than the light cylinder of most pulsars or magnetars.
This discovery provides convincing evidence that FRB emission originates from the complex magnetosphere of a magnetar, suggesting an FRB emission mechanism that is analogous to radio pulsars despite a huge luminosity difference between two types of objects.
\end{abstract}



\section{Introduction}\label{sec:intro}
Fast radio bursts (FRBs) are bright millisecond-duration astronomical transients predominantly originating from cosmological distances \citep{lorimer2007bright,thornton13,kirsten2022repeating}.
A robust FRB-magnetar connection was established with the detection of FRB 20200428D from a Galactic magnetar SGR\,J1935+2154 \citep{Bochenek20,CHIMEJ1935}. However, the origin(s) of cosmological FRBs are still subject to debate, and largely unknown \citep{ZhangB2023}. Within the framework of the magnetar engine, the emission site is not identified.

While pulsar-like mechanisms invoking magnetospheric emission are widely discussed, mechanisms invoking relativistic shocks far away from the magnetospheres are also discussed in the literature. Observationally, some pulsar-like emission behaviors have been reported via polarization measurements, including diverse polarization angle swings \citep{luo2020diverse}, S-shaped PA evolution \citep{mckinven2024pulsar}, and highly circular polarization \citep{xu2022fast,jiang2022fast,jiang2024ninetypercentcircularpolarization}. These observations add support to the magnetospheric origin for at least some FRBs. However, attempts to interpret these phenomena within the far-way models have also been proposed \citep{QuZhang2023,Iwamoto2024}. Smoking gun signatures to firmly establish the magnetospheric origin of FRB emission is called for.

Orthogonal jumps, that is, a sudden change in dominance between the two orthogonally polarised modes (OPMs), have been widely observed in pulsars \citep{manchester1975observations,stinebring1984pulsar,karastergiou2009complex}.
The OPMs indicate a strongly magnetized plasma environment in the magnetosphere of a pulsar \citep{arons1986wave,petrova2001origin}.
Birefringence can separate the two modes in the magnetospheric plasma \citep{1997ApJ...475..763M}, or the two modes are emitted from different regions that be seen sequentially due to the rotation, so that orthogonal jumps can be seen in the time domain.

In this Letter, we report the surprising sudden PA jumps recently discovered for the first time in a few bursts of FRB 20201124A, which is a repeater with high burst rate active windows. The FRB source was discovered by the CHIME \citep{Chime2021ATel14497} and subsequently located by the giant meter-wave radio telescope (uGMRT) \citep{Wharton2021ATel14538} and the European VLBI Network (EVN) \citep{Nimmo2021arXiv211101600N}. 
The Letter is organized as follows: In Section 2, we describe the observations and data processing process. In Section 3, we present the results of the PA jump of three bursts of FRB 20201124A and compare them with the cases of several pulsars. In Section 4, we show a simulation result and discuss theoretical models that may account for such jumps. The results are summarized in Section 5.

\section{OBSERVATIONS} \label{sec:style}
Observational data used in this Letter were taken using the Five-hundred-meter Aperture Spherical Telescope (FAST) telescope with its 19 beam receivers in 1.0-1.5\,GHz  \citep{nan2011five,li2018fast, jiang2019commissioning}. The data were recorded with the re-configurable open architecture computing hardware version 2 (ROACH2) System  \citep{hickish2016decade}. The recorded data are in the PSRFITS format with four polarizations, 49.152\,$\mu$s sampling interval, and 4096 channels in search mode. 
Short observations of a pulsed noise source were conducted by injecting it into the feed before the observation for calibration. We use the PSR/IEEE convention \citep{van2010psrchive} for the definition of the Stokes parameters and the \texttt{DSPSR}\footnote{\url{http://dspsr.sourceforge.net/}}  \citep{van2011dspsr} program and the \texttt{PSRCHIVE}\footnote{\url{http://psrchive.sourceforge.net/}}  \citep{van2012psrchive} software package for polarization analysis.

\begin{figure*}
  \centering
  \includegraphics[width=6.0in]{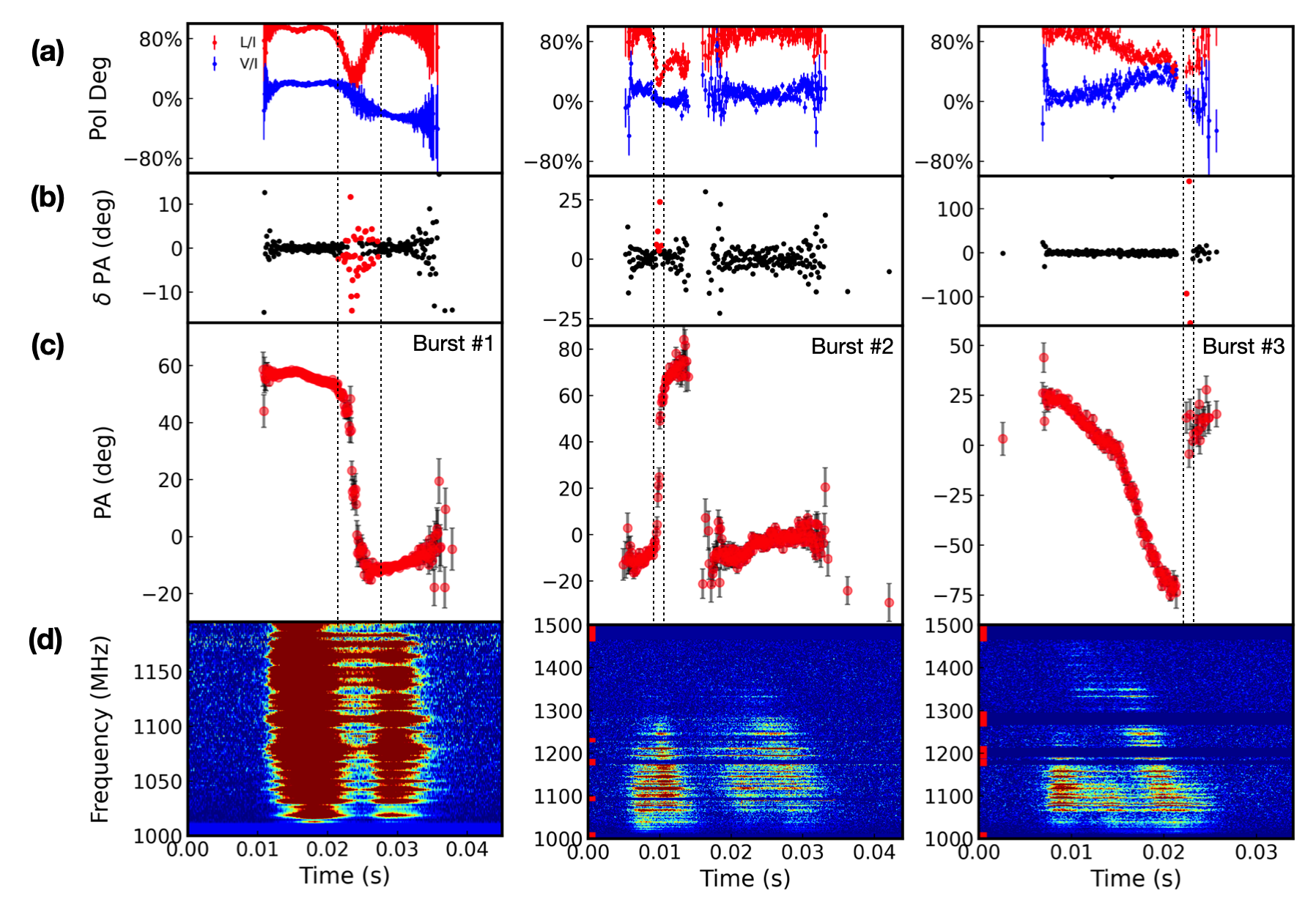}
  \caption{The PA orthogonal jump of three FRB\,20201124A bursts. Panel \textbf{(a)} shows the degrees of linear and circular polarization. Panel \textbf{(b)} shows the bin-to-bin difference of the PA curve. The red error bars in the panel \textbf{(c)} represent the measured PA swing of the three FRB bursts. The panel \textbf{(d)} shows the intensity of the three bursts in time and frequency. For Burst \#1 we show only the burst in a frequency range of 1000-1200\,MHz. For Burst \#2 and Burst \#3 we show the full 1000-1500\,MHz band. Using the outliers in the bin-to-bin difference of PA (marked as red dots in panel \textbf{(b)}), we find the time range in which the PA jumps occur, and mark them with the dotted vertical line in panel \textbf{(a)}, \textbf{(b)} and \textbf{(c)}.
}
  \label{fig:frborthogonal}
 \end{figure*}

In the first episode from 1 April 2021 to 11 June 2021 (UT), FAST identified 1,863 bursts with a signal-to-noise ratio higher than 7 \citep{xu2022fast}.
FAST observed the second episode on September 25-28, 2021, and they reported more than 600 bursts \citep{zhou2022fast,zhang2022fast,jiang2022fast,niu2022fast}. The maximum detection rate is more than 400 per hour. The degree of change in the PA curve is measured according to the maximum value of the PA change (defined in \cite{jiang2022fast}). Among all the PA variation measurable bursts, the flat PA cases (PA change is less than 10$^\circ$) accounted for $\sim 93\%$, and PA swing cases (PA change is larger than 10$^\circ$) accounted for $\sim 7\%$.

\begin{figure*}
  \centering
  \includegraphics[width=4.5in]{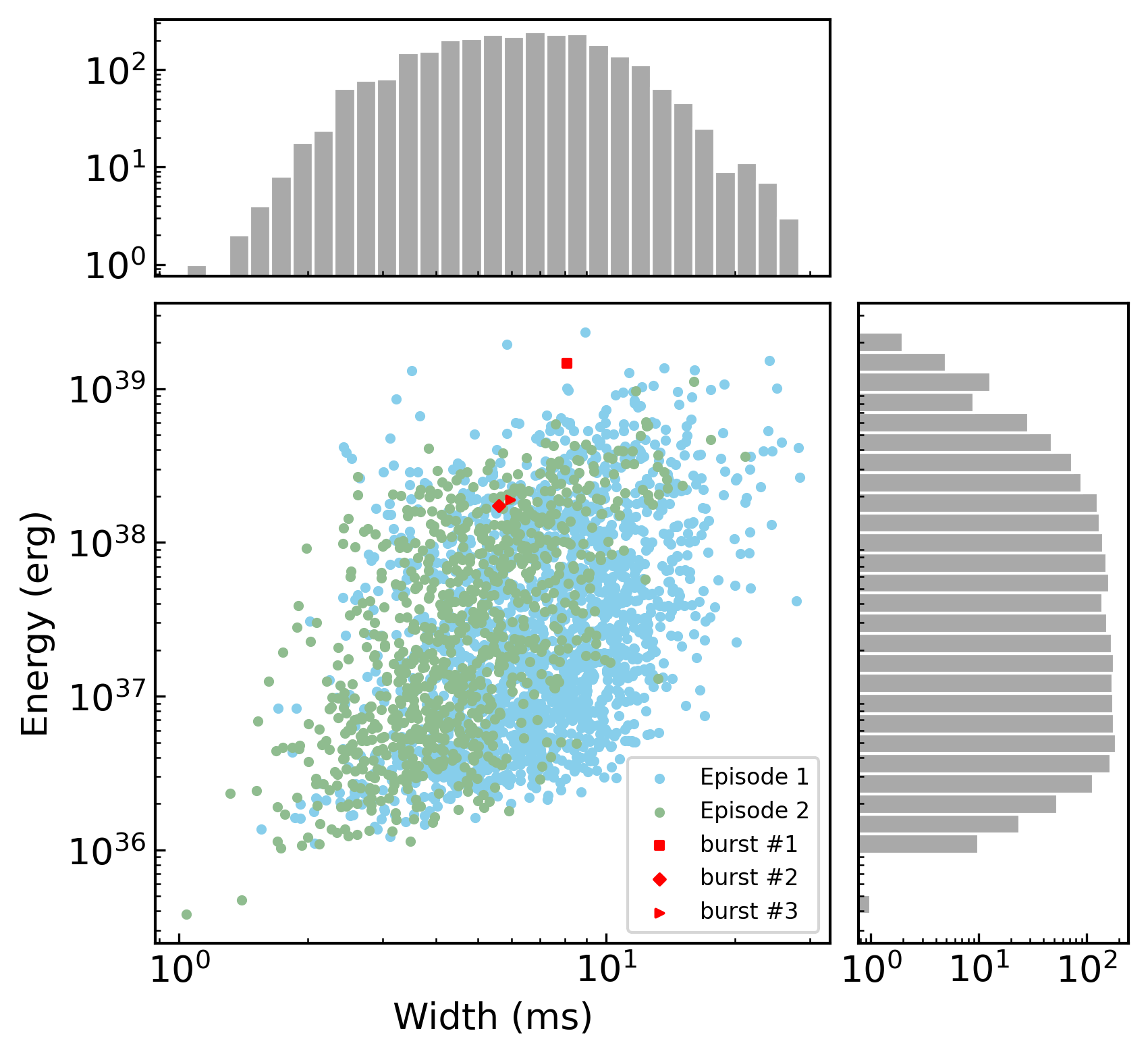}
  \caption{Burst width and energy distribution of FRB\,20201124A in the first and second episodes. We mark the OPM-jump bursts in red scatter. 
}
  \label{fig:energy}
 \end{figure*}

\begin{figure*}
  \centering
  \includegraphics[width=6.0in]{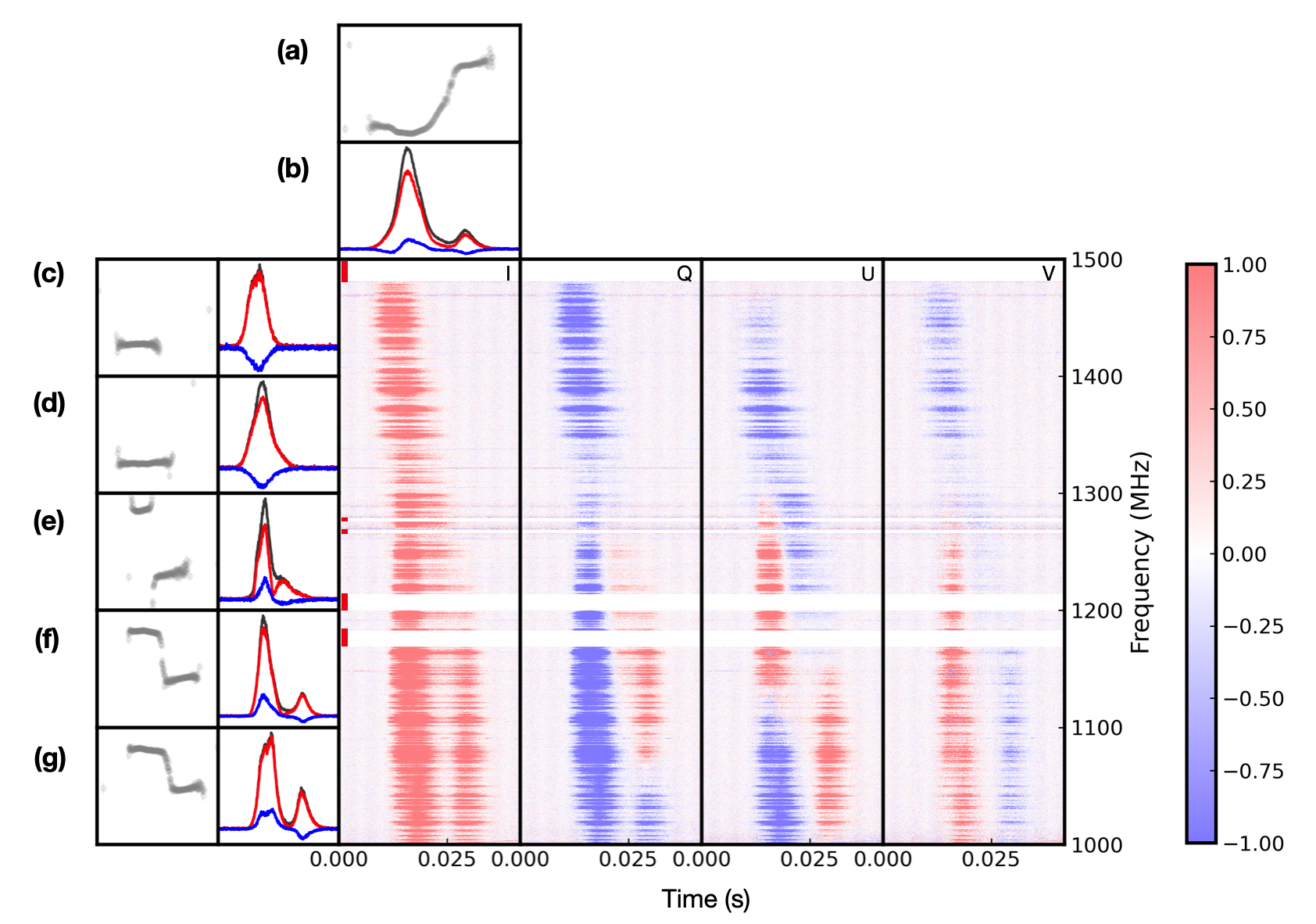}
  \caption{Polarization profiles of FRB\,20201124A burst \#1. The PA swing and the profile of the 1000-1500\,MHz band are shown in panels \textbf{(a)} and \textbf{(b)}. The polarization profile and PA of five 100\,MHz subbands are displayed separately on the left side of panel \textbf{(c)}, \textbf{(d)}, \textbf{(e)}, \textbf{(f)} and \textbf{(g)}. The black curve represents the total intensity $I$, the red linearly polarised intensity $L$, and the blue curve is circularly polarization $V$. The value of PA ranges from -90 to 90$^\circ$. The burst polarization evolves with time and frequency. On the right side of panel \textbf{(c)}, \textbf{(d)}, \textbf{(e)}, \textbf{(f)} and \textbf{(g)}, we show how the two-dimensional $I$,$Q$,$U$,$V$ varies with the horizontal axis --- time and the vertical axis --- frequency.}
  \label{fig:burst1all}
 \end{figure*}

\begin{table}
\caption{Information of PA jump bursts.} 
\label{tab:info}\setlength{\tabcolsep}{5.5mm}{
\begin{threeparttable}
\begin{tabular}{lllll}
\hline
Number & TOA (BC) \tnote{1}      & RM (rad m$^{-2}$)  & DM (pc cm$^{-3}$)                   & Fluence ($\rm mJy\,ms$)              \\
\hline
Burst\#1 & 59314.337398370  & -654.9 $\pm$ 0.8  & 413.5 $\pm$ 0.7  & 13554 $\pm$ 10  \\
Burst\#2 & 59485.808225740 & -610.9 $\pm$ 0.4  & 410.2 $\pm$ 1.8  & 2588 $\pm$ 26    \\
Burst\#3 & 59485.819949776 & -565.1 $\pm$ 2.6  & 411.8 $\pm$ 3.7  & 1698  $\pm$ 22  \\
\hline
\end{tabular}
\begin{tablenotes}
       \footnotesize
       \item[1] Barycentrical arrival time at 1500 MHz.
     \end{tablenotes}
    \end{threeparttable}}
\end{table}

We detected Burst \#1 with orthogonal PA jump in the first episode and Bursts \#2  and \#3 in the second episode, as shown in Figure \ref{fig:frborthogonal}. The measured arrival times, dispersion measure (DM) values, rotation measure (RM) values and fluence values are listed in Table \ref{tab:info}. The DM was determined based on the most prominent structure, and the fitted sigma was adopted as the DM error.

For comparison, We also presented the single-pulse and integrated polarization profiles from the pulsars PSR J1136+1551 (B1133+16) and PSR J0358+5413 (B0355+54) in Figure  \ref{fig:frbandpsr}. We folded the pulsars using the ephemerides obtained from the \texttt{PSRCAT}\footnote{\url{https://www.atnf.csiro.au/research/pulsar/psrcat/}} and obtained their integrated profiles. The rotation measure of PSR J1136+1551 is RM = 7.26 $\pm$ 0.02 rad m$^{-2}$ and that of PSR J0358+5413 is RM = 82.257 $\pm$ 0.002 rad m$^{-2}$.

To determine the time range of the PA jump of the FRB, we calculate the PA difference between adjacent time bins with the times bins in the PA curve uniformly sampled. When a jump occurs, the bin-to-bin difference of the PA curve  becomes significantly larger. We take all the points where the discrete difference is greater than 1 $\sigma$ of the baseline fluctuation to determine the jump range. After the jump range is determined, the mean values of PA before and after the jump are taken to calculate the jump angle. We use {PA}$_{err}$ as the weight to calculate the standard deviation and finally take the standard deviation through error propagation as the error of the PA jump.

As shown in Figure \ref{fig:energy}, we plot the energy and width of three OPM-jump bursts with respect to those of all the bursts detected in the two active episodes. We get the isotropic-equivalent burst energy 
\begin{equation} 
\begin{aligned} 
    E=\frac{4\pi d_L^2 F_\nu \Delta \nu}{1+z},
\end{aligned} 
\end{equation} 
where $d_L$ is the luminosity distance, $F_\nu$ is the specific fluence, $\Delta \nu$ is the bandwidth and $z$ is the redshift. Fluence and bandwidth information are from data tables published for the first and second active episodes \citep{xu2022fast,zhang2022fast}. The energy of Burst \#1 is 1.47 $\times$ 10${^{39}}$ erg. Burst \#2 and Burst \#3 have somewhat smaller energies of 1.74$\times$ 10$^{38}$ erg and 1.90 $\times$10$^{38}$ erg, respectively.

\section{Results} \label{sec:floats}
Most FRB events in our sample exhibit a high degree of polarization and a flat PA. Some bursts exhibit continuous PA swings with diverse patterns.
We present three bursts with polarization position angle jumps from FAST observations of FRB 20201124A, as shown in Figure \ref{fig:frborthogonal}. Burst \#1 shows a $65.9 \pm 1.1 $ degree jump in the 1000-1200~MHZ frequency in 4.1 ms. Burst \#2 shows a $97.7 \pm 23.9 $ degree jump in 1.2 ms. 
Burst \#3 shows a $88.7 \pm 30.0 $ degree jump in 1.4 ms. The PA jumps of Bursts \#2 and \#3 can be regarded as 90-degree jumps within the error range, while the jump of Burst \#1 is smaller than 90$^\circ$.
Such jumps in the PAs have been observed in many pulsars
\citep{manchester1975observations,cordes1978orthogonal,backer1980statistical,stinebring1984pulsar,karastergiou2009complex}.

\begin{figure*}
  \centering
  \includegraphics[width=6.3in]{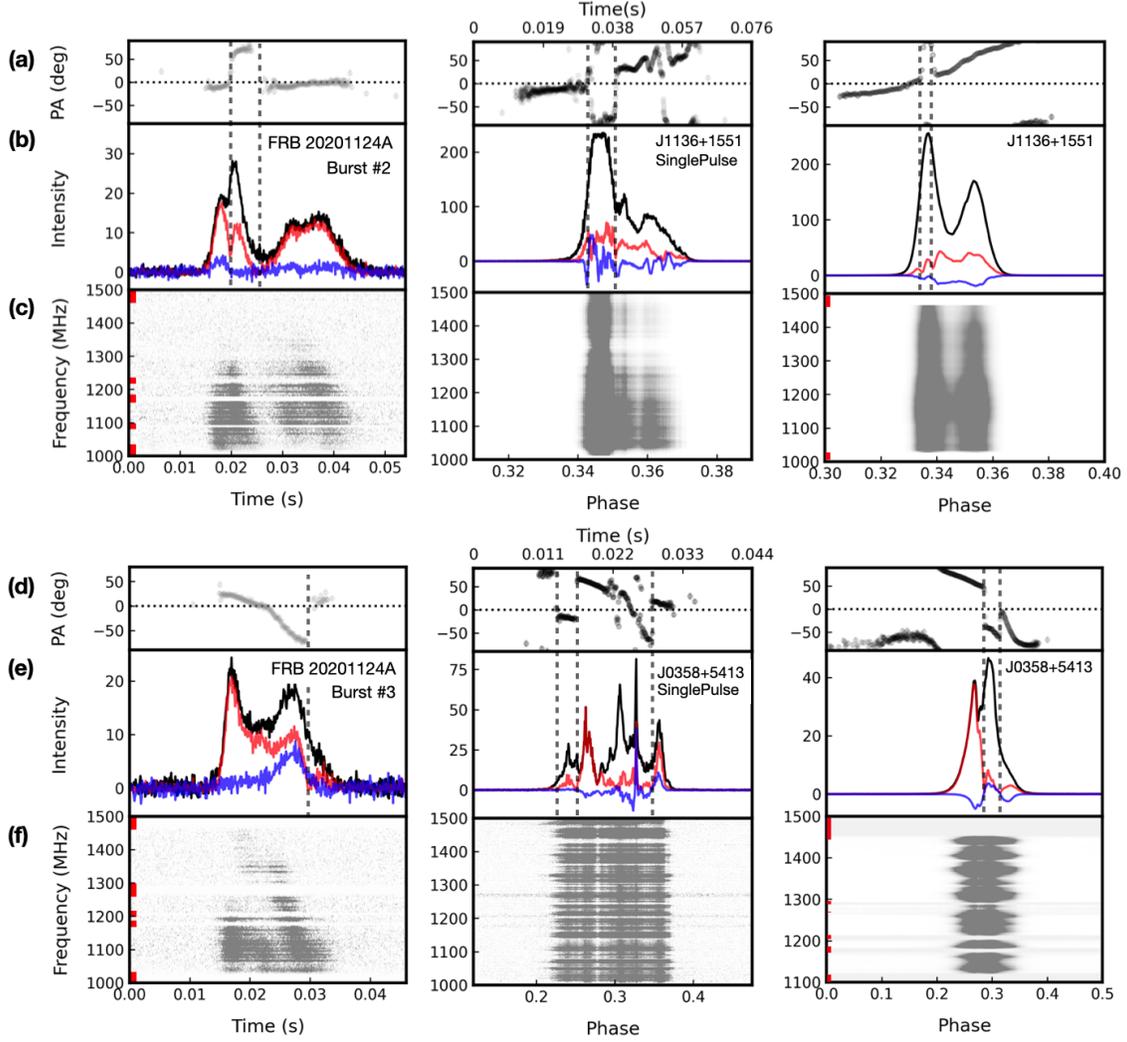}
  \caption{Comparing FRB\,20201124A OPM-jump bursts \#2 and \#3 with the single pulse and integrated profile of Pulsar J0358+5413 and J1136+1551. In panel \textbf{(a)}, \textbf{(b)} and \textbf{(c)}, we show bursts \#2 and PSR J1136+1551's single pulse and integrated PA, polarization profile and waterfall, respectively.
In panel \textbf{(d)}, \textbf{(e)} and \textbf{(f)}, we show the same information for bursts \#3 and PSR J0358+5413. The dotted horizontal lines show the 0$^\circ$ PA. The polarization profiles are shown in the panel \textbf{(b)} and \textbf{(e)}, with linear polarization in red, circular polarization in blue, and total intensity in black. The intensity of the profiles is normalized such that the off-pulse root mean square equals unity. The positions of the dotted vertical lines indicate the time (phase) of the orthogonal jumps.}
  \label{fig:frbandpsr}
 \end{figure*}

Unlike Bursts \#2 and \#3, the PA jump of Burst \#1 is not 90 degrees. Its burst morphology is more complex and composed of multiple components.
Figure \ref{fig:burst1all} panel B shows that its 1000 MHz--1500 MHz profile has a double-peak structure with a PA swing. 
In Figure \ref{fig:burst1all}, we plot the $I$, $Q$, $U$, $V$ components of the burst as a function of frequency after polarization calibration and de-rotation. Normally, $Q$, $U$, and $V$ should be the same across frequencies for any given time if the PA angles are identical in different frequencies. However, the Stokes parameters of the same time sample change significantly in frequency with sign changes. This suggests that there are frequency-dependent polarization components that should be treated separately. 
So, we divide the 1000--1500\,MHz into five sub-bands and show the profile and PA of the sub-bands, respectively. 
The 1300-1400\ MHz and 1400--1500\ MHz profiles show a single peak with negative circular polarization and flat PA. The 1000--1100\,MHz, 1100--1200\,MHz, and 1200-1300\,MHz profiles show double peaks and an orthogonal jump of PA between the peaks. 
The double peak in low-frequency results from the emergence of a sub-burst component $\sim5$\,ms after the main peak, as part of a typical FRB down-drifting pattern.

The FRB PA jumps are analogous to those of pulsars' single pulses.
For comparison, we select two bright pulsars observed by FAST and analyze their polarization properties.
In Figure \ref{fig:frbandpsr}, we compare the profiles of Bursts \#2 and \#3 from FRB\,20201124A with the single-pulse and integrated profiles from PSR~J1136+1551 and PSR~J0358+5413. Pulsars and FRBs show some similarities in polarization properties.
The PA swings observed in the pulsar single pulse and integrated profiles show clear OPMs.
The OPM jump could happen at different times in the bursts. For instance, it could be in the middle of the PA swing like FRB\,20201124A Burst \#2 and PSR~J1136+1551, or it could be in the tail like FRB\,20201124A Burst \#3 and PSR~J0358+5413.
The FRB jumps happen where the linear polarization decreases or circular polarization changes sign, similar to what is observed in pulsars \citep{manchester1975observations,radhakrishnan1990toward,xilouris1995pulsar,2024MNRAS.527.2612S}.
The sign change of circular polarization is attributed to the OPMs with different orientations of circular polarization.
Furthermore, in pulsars, the jumps do not have to be exactly 90$^\circ$.
The PA jump of Burst \#1 (Figure \ref{fig:frborthogonal}) is about 66$^\circ$.
Some pulsar OPM studies have also found evidence of deviations from pure orthogonality, with the PA swing not separated by precisely 90$^\circ$.
This could be explained by the superposition of non-orthogonal modes originating from different field lines and birefringent refraction \citep{stinebring1984pulsar,mckinnon2003transition}.

The three bursts with OPM jumps were found among the $\sim$2500 bursts collected from two active episodes of FRB 20201124A. The isotropic energies, as shown in Figure \ref{fig:energy}, are among the luminous ones. The highest-energy Burst \#1 has an energy exceeding 10$^{39}$ erg, which is larger than most other bursts. 
FRB bursts are generally far more energetic than pulsar pulses. 
Theoretically, it is possible that low-energy FRBs may be generated with mechanisms more akin to pulsar emission. However, the fact that the OPM-jump bursts belong to the bright sample of FRB bursts disfavors the view that only fainter FRBs are more analogous to pulsars. It suggests a possible common mechanism of radio emission despite the huge energy difference between FRBs and pulsars. The fact that OPM-jump bursts are quite rare indicates that very special conditions, likely special geometric configurations, are required to cause the OPM-jump phenomenon.

\section{Physical origin}
\subsection{Superposition of multi-plane waves}
\begin{figure*}
  \centering
  \includegraphics[width=4.5in]{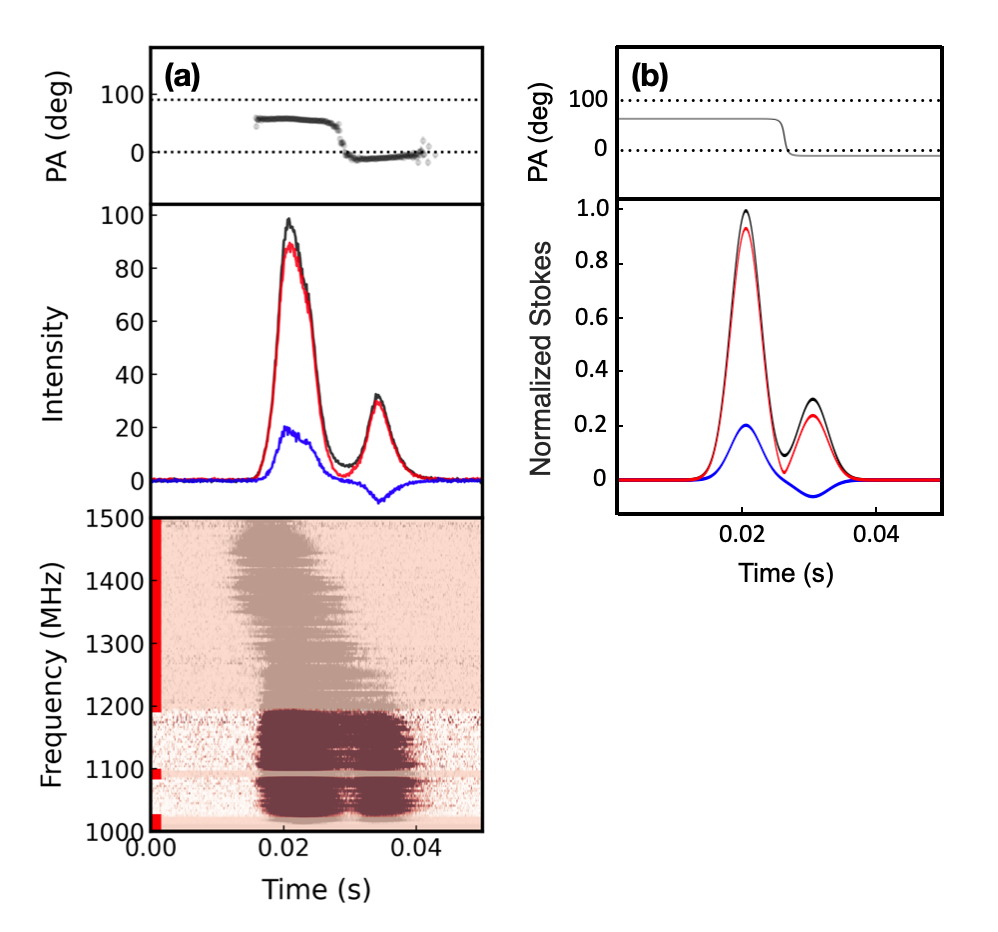}
  \caption{Observed and Simulated Stokes profile FRB\,20201124A Burst \#1. Burst \#1 in frequency bands 1021-1190\,MHz with DM of 413.5 pc\,cm$^{-3}$ shows in panel \textbf{(a)}. The polarization position angle of the low-frequency double-peak structure exhibits an orthogonal jump phenomenon with a jump angle of 66$^\circ$. The panel \textbf{(b)} shows the simulated Stokes profile and its corresponding PA swing.}
  \label{fig:sim}
 \end{figure*}

One possible mechanism to explain the orthogonal jump is the incoherent superposition by multi-plane waves.
When the components correspond to different orthogonal modes in an incoherent superposition, the orthogonal modes can occur at the point of minimum linear polarization \citep{2024MNRAS.527.2612S}.
To explain Burst \#1, we simulate the incoherent superposition of two components, each consisting of only Gaussian Stokes profiles (Figure \ref{fig:sim}). 
The Stokes profiles are given by 
\begin{equation} 
\begin{aligned} 
&I=\exp(-[(t-0.02)/\sigma_w]^2)+0.3\exp(-[(t-0.03)/\sigma_w]^2),\\ 
&Q=-0.4\exp(-[(t-0.02)/\sigma_w]^2)+0.2\exp(-[(t-0.03)/\sigma_w]^2),\\ 
&U=-0.8\exp(-[(t-0.02)/\sigma_w]^2)-0.08\exp(-[(t-0.03)/\sigma_w]^2),\\ 
&V=0.2\exp(-[(t-0.02)/\sigma_w]^2)-0.06\exp(-[(t-0.03)/\sigma_w]^2), 
\end{aligned} 
\end{equation} 
where $t$ is in unit of second and $\sigma_w=3.2\times10^{-3}$ s.
The corresponding linear polarization fraction and PA are given by $L=\sqrt{U^2+Q^2}$ and $\psi=1/2\tan^{-1}(U/Q)$.
For simplicity, each component has a polarization degree close to $100\%$.
One can see that the superposition of two simple components could reproduce the main features of orthogonal jumps, i.e., the PA jump of $\sim66^\circ$ and the linear and circular polarization changes observed in the most complicated case, Burst \#1.
The PA jump can occur at minimum linear polarization.
Because the PA is the angle with the origin of the $Q-U$ coordinate as the vertex, it generally has a discontinuity, i.e., a sudden jump, when the line of sight travels across a point of $L=0$ in the Poincar\'e sphere \citep{1997A&A...323..395X}.
A jump with $\delta \Psi<90^\circ$ can occur at a minimum $L$ with a small number but not strictly zero.

\subsection{Emission mechanisms and propagation effects}

In the highly magnetized magnetosphere of magnetars, the two orthogonal modes for which the position angles differ by $90^\circ$ are reminiscent of the O-mode and X-mode waves \citep{1986ApJ...302..138B}: the O-mode is polarized in the plane defined by the background magnetic field and the wave vector, i.e. $\vec E_w\parallel(\vec k-\vec B_{\rm bg})$ and the X-mode is polarized perpendicularly to the magnetic field and the wave vector, i.e. $\vec E_w\perp(\vec k-\vec B_{\rm bg})$. In the context of FRBs, their emission mechanism still remains a mystery. There are two classes of models depending on whether the FRBs are created inside the magnetar magnetosphere or far away from the magnetosphere. Commonly discussed magnetospheric models invoke photons generated via curvature radiation \citep{katz2014,2017MNRAS.468.2726K,2018ApJ...868...31Y,2020MNRAS.498.1397L,Kumar2020,2020ApJ...899..109W,2021MNRAS.508L..32C,2022ApJ...927..105W} and inverse Compton scattering off low-frequency fast magnetosonic waves \citep{2022ApJ...925...53Z,QuZhang2023,2024arXiv240411948Q}. Both mechanisms can generate X-mode and O-mode at $\sim$GHz bands and these modes can propagate freely within the magnetosphere \citep{2022MNRAS.515.2020Q}. 
Within these models, FRBs are likely produced via a series of charged bunches and different emission patches could contribute to the incoherent superposition of the two modes. 

We emphasize that a sudden change of the dominant mode over observing time can be realized only within the magnetosphere of a rotating neutron star. The dominance of two modes before and after the jump time is more naturally interpreted by assuming that the two dominant emission modes come from two distinct emission regions. This can be realized for a rotating magnetosphere with different observing times corresponding to different rotation phases as the line of sight sweeps across different field lines \citep{cheng1979theory,wang2014polarized}. The OPM jump time corresponds to the epoch when the dominance of the two modes switches.

Outside the magnetosphere, synchrotron maser emission may occur when the relativistic outflow interacts with a highly magnetized plasma to form a quasi-perpendicular shock. 
The PA of the synchrotron maser emission can also vary due to the generation of the O-mode waves \citep[e.g.][]{Iwamoto2024}. 
However, there are several issues with such a mechanism: 
First, in order to produce the observed jumps, the relative amplitude of the two orthogonal modes should be small. However, within the synchrotron maser model, the X-mode wave amplitude is usually much larger than that of the O-mode wave due to the effect of Alf\'ven ion cyclotron instability on O-mode waves \citep{Iwamoto2024}. A small magnetization is required to generate O mode with a comparable amplitude as X mode, but this would lead to a relatively low linear polarization degree. Second and more severely, unlike the pulsar-like model that can interpret a rapid switch between two orthogonal modes via line of sight sweeping across different emission regions, the far-away shock models suggest that the same emitting plasma is continuously observed as a function of time. There is no known mechanism that can cause a sudden change in the dominant polarization mode at a certain time. One contrived scenario would be to assume that the emission comes from small patches of plasma with a size much smaller than the causally connected region and that such emitting regions fluctuate with time. However, this would decrease the radiation efficiency by a significant factor \citep{lu2020unified}, and no far-away coherent mechanism is known to achieve this.

It is worth pointing out that both coherent and incoherent superpositions of multiple wave modes are possible.
For a coherent superposition of X-mode and O-mode, the total polarization degree is always 100\%.
For an incoherent superposition, on the other hand, significant depolarization can occur. The latter matches the observation of FRB 20201124A (see Figure \ref{fig:sim}). For both coherent and incoherent superposition cases, a PA jump occurs when the linear polarization reaches zero, but the difference is that $|V|$ must have the largest value only for the coherent superposition.
The photons from the two modes may be produced from different emission regions, which may lead to incoherent superposition to cause the PA jump. However, it is not easy to identify which sub-pulse is O(X)-mode due to the unknown magnetic orientation and charge sign of emitting particles.

Finally, plasma lensing may in principle produce OPM jumps \citep[e.g.][]{Er23}. However, in order to satisfy the observational constraints such as millisecond OPM jump timescale and the probability of detecting a jump during several bursts, contrived model parameters have to be invoked.

\subsection{OPM Jump timescale constraint} 

One can also pose a generic constraint of the emission region based on the timescale of the observed OPM jumps. The very short $\sim1$ ms PA jump timescale corresponds to a light-crossing distance of $\sim 300$ km. Such a scale is smaller than the light cylinder radius of a typical magnetar. Without getting to the details of emission models, this timescale again hints at an emission region inside a neutron star magnetosphere. 

Under the assumption of magnetospheric emission, one can use the OPM jump timescale to place a very loose lower limit of the magnetar spin period. The jump time of $\sim 4.1$ ms places the spin period to be longer than this, which excludes a millisecond magnetar as the source of this FRB. This is as expected, because a magnetar with such a short period has a spindown timescale (the characteristic timescale defined by $-\Omega/\dot\Omega$, where $\Omega$ is the angular frequency and $\dot\Omega$ is its decreasing rate) of $t_{\rm sd} = 3c^3 I/(B^2R^6 \Omega^2) \simeq (0.4 {\rm d}) I_{45} B_{p,15}^{-2} (P/4.1~{\rm ms})^2 R_6^{-6}$ (where $B_p=(10^{15} ~{\rm G}) B_{p,15}$ is the dipolar magnetic field strength at the magnetic pole, $\Omega=2\pi/P$ is the angular velocity, $I=(10^{45}~{\rm g~cm^2}) I_{45}$ is the moment of inertia, and $R=10^6 ~{\rm cm}R_6$ is the radius of the neutron star). This is much shorter than the active period of FRB 20201124A, which is years.

\section{Summary}

In this Letter, we reported three bright bursts from FRB 20201124A that show significant OPM jumps. We showed that such jumps are analogous to the phenomenon commonly observed in pulsars. This suggests that the FRB mechanism is pulsar-like, and should originate from the rotating magnetopshere of a central engine, likely a young magnetar. Our results suggest that even the bright bursts of repeating FRBs are generated within the magnetospheres and the far-away synchrotron maser shock models are strongly disfavored.

\section{Acknowledgments}
\begin{acknowledgments}
We thank the anonymous referee for insightful comments and suggestions that significantly improved our manuscript.
This work made use of data from the FAST, a Chinese national mega-science facility, built and operated by the National Astronomical Observatories, Chinese Academy of Sciences. 
The authors acknowledge support from the National SKA Program of China No. 2020SKA0120200, 2020SKA0120100 (W.W.Zhu, K.J.Lee, H.Xu, D.J.Zhou, J.R.Niu, J.C.Jiang), 
the National Nature Science Foundation grant No. 12041303 (W.W.Zhu, P.Wang),  12041304(Y.L.Yue)
CAS Project for Young Scientists in Basic Research, YSBR-063 (W.W.Zhu, P.Jiang, C.H.Niu, X.L.Miao),
the National Key R\&D Program of China No. 2017YFA0402600, 2021YFA0718500 (P.Wang, D.Li, Y.L.Yue, Y.K.Zhang)
the CAS-MPG LEGACY project, funding from the Max-Planck Partner Group (P.Wang, D.Li, Y.L.Yue, K.J.Lee), the Strategic Priority Research Program of the Chinese Academy of Sciences, Grant No. XDB0550300 (W.Y.Wang).
and the Leading Innovation and Entrepreneurship Team of Zhejiang Province of China Grant No. 2023R01008(Y. Feng).
\end{acknowledgments}

\section{Author Contributions}
J.R. Niu, W.Y. Wang, and J.C. Jiang contributed equally and led the data analysis and paper writing. W.W. Zhu, K.J. Lee, J.L. Han, and B. Zhang coordinated the radio observational campaign, science investigation and paper writing. All authors participated in discussions and the writing of the paper.

\bibliography{sample631}{}

\begin{thebibliography}{}
\expandafter\ifx\csname natexlab\endcsname\relax\def\natexlab#1{#1}\fi
\providecommand{\url}[1]{\href{#1}{#1}}
\providecommand{\dodoi}[1]{doi:~\href{http://doi.org/#1}{\nolinkurl{#1}}}
\providecommand{\doeprint}[1]{\href{http://ascl.net/#1}{\nolinkurl{http://ascl.net/#1}}}
\providecommand{\doarXiv}[1]{\href{https://arxiv.org/abs/#1}{\nolinkurl{https://arxiv.org/abs/#1}}}

\bibitem[{Arons \& Barnard(1986)}]{arons1986wave}
Arons, J., \& Barnard, J.~J. 1986, The Astrophysical Journal, 302, 120

\bibitem[{Backer \& Rankin(1980)}]{backer1980statistical}
Backer, D., \& Rankin, J.~M. 1980, The Astrophysical Journal Supplement Series,
  42, 143

\bibitem[{Barnard \& Arons(1986)}]{1986ApJ...302..138B}
Barnard, J.~J., \& Arons, J. 1986, Astrophysical Journal, Part 1 (ISSN
  0004-637X), vol. 302, March 1, 1986, p. 138-162. Research supported by the
  University of California., 302, 138

\bibitem[{Bochenek {et~al.}(2020)Bochenek, Ravi, Belov, Hallinan, Kocz,
  Kulkarni, \& McKenna}]{Bochenek20}
Bochenek, C.~D., Ravi, V., Belov, K.~V., {et~al.} 2020, Nature, 587, 59

\bibitem[{Cheng \& Ruderman(1979)}]{cheng1979theory}
Cheng, A.~F., \& Ruderman, M. 1979, Astrophysical Journal, Part 1, vol. 229,
  Apr. 1, 1979, p. 348-360. NSF-supported research., 229, 348

\bibitem[{{CHIME/FRB Collaboration} {et~al.}(2020){CHIME/FRB Collaboration},
  {Andersen}, {Bandura}, {Bhardwaj}, {Bij}, {Boyce}, {Boyle}, {Brar},
  {Cassanelli}, {Chawla}, {Chen}, {Cliche}, {Cook}, {Cubranic}, {Curtin},
  {Denman}, {Dobbs}, {Dong}, {Fandino}, {Fonseca}, {Gaensler}, {Giri}, {Good},
  {Halpern}, {Hill}, {Hinshaw}, {H{\"o}fer}, {Josephy}, {Kania}, {Kaspi},
  {Landecker}, {Leung}, {Li}, {Lin}, {Masui}, {McKinven}, {Mena-Parra},
  {Merryfield}, {Meyers}, {Michilli}, {Milutinovic}, {Mirhosseini},
  {M{\"u}nchmeyer}, {Naidu}, {Newburgh}, {Ng}, {Patel}, {Pen},
  {Pinsonneault-Marotte}, {Pleunis}, {Quine}, {Rafiei-Ravandi}, {Rahman},
  {Ransom}, {Renard}, {Sanghavi}, {Scholz}, {Shaw}, {Shin}, {Siegel}, {Singh},
  {Smegal}, {Smith}, {Stairs}, {Tan}, {Tendulkar}, {Tretyakov}, {Vanderlinde},
  {Wang}, {Wulf}, \& {Zwaniga}}]{CHIMEJ1935}
{CHIME/FRB Collaboration}, {Andersen}, B.~C., {Bandura}, K.~M., {et~al.} 2020,
  \nat, 587, 54, \dodoi{10.1038/s41586-020-2863-y}

\bibitem[{{Chime/FRB Collabortion}(2021)}]{Chime2021ATel14497}
{Chime/FRB Collabortion}. 2021, The Astronomer's Telegram, 14497, 1

\bibitem[{{Cooper} \& {Wijers}(2021)}]{2021MNRAS.508L..32C}
{Cooper}, A.~J., \& {Wijers}, R.~A.~M.~J. 2021, \mnras, 508, L32,
  \dodoi{10.1093/mnrasl/slab099}

\bibitem[{Cordes {et~al.}(1978)Cordes, Rankin, \&
  Backer}]{cordes1978orthogonal}
Cordes, J., Rankin, J., \& Backer, D. 1978, The Astrophysical Journal, 223, 961

\bibitem[{{Er} {et~al.}(2023){Er}, {Pen}, {Sun}, \& {Li}}]{Er23}
{Er}, X., {Pen}, U.-L., {Sun}, X., \& {Li}, D. 2023, \mnras, 522, 3965,
  \dodoi{10.1093/mnras/stad1282}

\bibitem[{Hickish {et~al.}(2016)Hickish, Abdurashidova, Ali, Buch, Chaudhari,
  Chen, Dexter, Domagalski, Ford, Foster, {et~al.}}]{hickish2016decade}
Hickish, J., Abdurashidova, Z., Ali, Z., {et~al.} 2016, Journal of Astronomical
  Instrumentation, 5, 1641001

\bibitem[{{Iwamoto} {et~al.}(2024){Iwamoto}, {Matsumoto}, {Amano}, {Matsukiyo},
  \& {Hoshino}}]{Iwamoto2024}
{Iwamoto}, M., {Matsumoto}, Y., {Amano}, T., {Matsukiyo}, S., \& {Hoshino}, M.
  2024, \prl, 132, 035201, \dodoi{10.1103/PhysRevLett.132.035201}

\bibitem[{Jiang {et~al.}(2022)Jiang, Wang, Xu, Xu, Zhang, Wang, Zhou, Zhang,
  Niu, Lee, {et~al.}}]{jiang2022fast}
Jiang, J.-C., Wang, W.-Y., Xu, H., {et~al.} 2022, Research in Astronomy and
  Astrophysics, 22, 124003

\bibitem[{Jiang {et~al.}(2024)Jiang, Xu, Niu, Lee, Zhu, Zhang, Qu, Xu, Zhou,
  Cao, Wang, Wang, Cao, Zhang, Zhang, Gan, Han, Hao, Huang, Jiang, Li, Li, Li,
  Li, Luo, Men, Qian, Sun, Wang, Xu, Xu, Yang, Yao, Yue, Yu, Yuan, \&
  Zhu}]{jiang2024ninetypercentcircularpolarization}
Jiang, J.~C., Xu, J.~W., Niu, J.~R., {et~al.} 2024, Ninety percent circular
  polarization detected in a repeating fast radio burst.
\newblock \doarXiv{2408.03313}

\bibitem[{Jiang {et~al.}(2019)Jiang, Yue, Gan, Yao, Li, Pan, Sun, Yu, Liu,
  Tang, {et~al.}}]{jiang2019commissioning}
Jiang, P., Yue, Y., Gan, H., {et~al.} 2019, arXiv preprint arXiv:1903.06324

\bibitem[{Karastergiou(2009)}]{karastergiou2009complex}
Karastergiou, A. 2009, Monthly Notices of the Royal Astronomical Society:
  Letters, 392, L60

\bibitem[{{Katz}(2014)}]{katz2014}
{Katz}, J.~I. 2014, \prd, 89, 103009, \dodoi{10.1103/PhysRevD.89.103009}

\bibitem[{Kirsten {et~al.}(2022)Kirsten, Marcote, Nimmo, Hessels, Bhardwaj,
  Tendulkar, Keimpema, Yang, Snelders, Scholz, {et~al.}}]{kirsten2022repeating}
Kirsten, F., Marcote, B., Nimmo, K., {et~al.} 2022, Nature, 602, 585

\bibitem[{{Kumar} \& {Bo{\v{s}}njak}(2020)}]{Kumar2020}
{Kumar}, P., \& {Bo{\v{s}}njak}, {\v{Z}}. 2020, \mnras, 494, 2385,
  \dodoi{10.1093/mnras/staa774}

\bibitem[{{Kumar} {et~al.}(2017){Kumar}, {Lu}, \&
  {Bhattacharya}}]{2017MNRAS.468.2726K}
{Kumar}, P., {Lu}, W., \& {Bhattacharya}, M. 2017, \mnras, 468, 2726,
  \dodoi{10.1093/mnras/stx665}

\bibitem[{Li {et~al.}(2018)Li, Wang, Qian, Krco, Dunning, Jiang, Yue, Jin, Zhu,
  Pan, {et~al.}}]{li2018fast}
Li, D., Wang, P., Qian, L., {et~al.} 2018, IEEE Microwave Magazine, 19, 112

\bibitem[{Lorimer {et~al.}(2007)Lorimer, Bailes, McLaughlin, Narkevic, \&
  Crawford}]{lorimer2007bright}
Lorimer, D.~R., Bailes, M., McLaughlin, M.~A., Narkevic, D.~J., \& Crawford, F.
  2007, Science, 318, 777

\bibitem[{{Lu} {et~al.}(2020){Lu}, {Kumar}, \& {Zhang}}]{2020MNRAS.498.1397L}
{Lu}, W., {Kumar}, P., \& {Zhang}, B. 2020, \mnras, 498, 1397,
  \dodoi{10.1093/mnras/staa2450}

\bibitem[{Lu {et~al.}(2020)Lu, Kumar, \& Zhang}]{lu2020unified}
Lu, W., Kumar, P., \& Zhang, B. 2020, Monthly Notices of the Royal Astronomical
  Society, 498, 1397

\bibitem[{Luo {et~al.}(2020)Luo, Wang, Men, Zhang, Jiang, Xu, Wang, Lee, Han,
  Zhang, {et~al.}}]{luo2020diverse}
Luo, R., Wang, B., Men, Y., {et~al.} 2020, Nature, 586, 693

\bibitem[{Manchester {et~al.}(1975)Manchester, Taylor, \&
  Huguenin}]{manchester1975observations}
Manchester, R., Taylor, J., \& Huguenin, G. 1975, Astrophysical Journal, vol.
  196, Feb. 15, 1975, pt. 1, p. 83-102., 196, 83

\bibitem[{{McKinnon}(1997)}]{1997ApJ...475..763M}
{McKinnon}, M.~M. 1997, \apj, 475, 763, \dodoi{10.1086/303542}

\bibitem[{McKinnon(2003)}]{mckinnon2003transition}
McKinnon, M.~M. 2003, The Astrophysical Journal, 590, 1026

\bibitem[{Mckinven {et~al.}(2024)Mckinven, Bhardwaj, Eftekhari, Kilpatrick,
  Kirichenko, Pal, Cook, Gaensler, Giri, Kaspi, {et~al.}}]{mckinven2024pulsar}
Mckinven, R., Bhardwaj, M., Eftekhari, T., {et~al.} 2024, arXiv preprint
  arXiv:2402.09304

\bibitem[{Nan {et~al.}(2011)Nan, Li, Jin, Wang, Zhu, Zhu, Zhang, Yue, \&
  Qian}]{nan2011five}
Nan, R., Li, D., Jin, C., {et~al.} 2011, International Journal of Modern
  Physics D, 20, 989

\bibitem[{{Nimmo} {et~al.}(2021){Nimmo}, {Hewitt}, {Hessels}, {Kirsten},
  {Marcote}, {Bach}, {Blaauw}, {Burgay}, {Corongiu}, {Feiler}, {Gawro{\'n}ski},
  {Giroletti}, {Karuppusamy}, {Keimpema}, {Kharinov}, {Lindqvist},
  {Maccaferri}, {Melnikov}, {Mikhailov}, {Ould-Boukattine}, {Paragi}, {Pilia},
  {Possenti}, {Snelders}, {Surcis}, {Trudu}, {Venturi}, {Vlemmings}, {Wang},
  {Yang}, \& {Yuan}}]{Nimmo2021arXiv211101600N}
{Nimmo}, K., {Hewitt}, D.~M., {Hessels}, J.~W.~T., {et~al.} 2021, arXiv
  e-prints, arXiv:2111.01600.
\newblock \doarXiv{2111.01600}

\bibitem[{Niu {et~al.}(2022)Niu, Zhu, Zhang, Yuan, Zhou, Zhang, Jiang, Han, Li,
  Lee, {et~al.}}]{niu2022fast}
Niu, J.-R., Zhu, W.-W., Zhang, B., {et~al.} 2022, Research in Astronomy and
  Astrophysics, 22, 124004

\bibitem[{Petrova(2001)}]{petrova2001origin}
Petrova, S. 2001, Astronomy \& Astrophysics, 378, 883

\bibitem[{{Qu} {et~al.}(2022){Qu}, {Kumar}, \& {Zhang}}]{2022MNRAS.515.2020Q}
{Qu}, Y., {Kumar}, P., \& {Zhang}, B. 2022, \mnras, 515, 2020,
  \dodoi{10.1093/mnras/stac1910}

\bibitem[{Qu \& Zhang(2023)}]{QuZhang2023}
Qu, Y., \& Zhang, B. 2023, Monthly Notices of the Royal Astronomical Society,
  522, 2448

\bibitem[{{Qu} \& {Zhang}(2024)}]{2024arXiv240411948Q}
{Qu}, Y., \& {Zhang}, B. 2024, arXiv e-prints, arXiv:2404.11948,
  \dodoi{10.48550/arXiv.2404.11948}

\bibitem[{Radhakrishnan \& Rankin(1990)}]{radhakrishnan1990toward}
Radhakrishnan, V., \& Rankin, J.~M. 1990

\bibitem[{{Singh} {et~al.}(2024){Singh}, {Gupta}, \&
  {De}}]{2024MNRAS.527.2612S}
{Singh}, S., {Gupta}, Y., \& {De}, K. 2024, \mnras, 527, 2612,
  \dodoi{10.1093/mnras/stad3334}

\bibitem[{Stinebring {et~al.}(1984)Stinebring, Cordes, Rankin, Weisberg, \&
  Boriakoff}]{stinebring1984pulsar}
Stinebring, D.~R., Cordes, J., Rankin, J.~M., Weisberg, J., \& Boriakoff, V.
  1984, The Astrophysical Journal Supplement Series, 55, 247

\bibitem[{{Thornton} {et~al.}(2013){Thornton}, {Stappers}, {Bailes},
  {Barsdell}, {Bates}, {Bhat}, {Burgay}, {Burke-Spolaor}, {Champion}, {Coster},
  {D'Amico}, {Jameson}, {Johnston}, {Keith}, {Kramer}, {Levin}, {Milia}, {Ng},
  {Possenti}, \& {van Straten}}]{thornton13}
{Thornton}, D., {Stappers}, B., {Bailes}, M., {et~al.} 2013, Science, 341, 53,
  \dodoi{10.1126/science.1236789}

\bibitem[{van Straten \& Bailes(2011)}]{van2011dspsr}
van Straten, W., \& Bailes, M. 2011, Publications of the Astronomical Society
  of Australia, 28, 1

\bibitem[{van Straten {et~al.}(2012)van Straten, Demorest, \&
  Os{\l}owski}]{van2012psrchive}
van Straten, W., Demorest, P., \& Os{\l}owski, S. 2012, arXiv preprint
  arXiv:1205.6276

\bibitem[{Van~Straten {et~al.}(2010)Van~Straten, Manchester, Johnston, \&
  Reynolds}]{van2010psrchive}
Van~Straten, W., Manchester, R., Johnston, S., \& Reynolds, J. 2010,
  Publications of the Astronomical Society of Australia, 27, 104

\bibitem[{Wang {et~al.}(2014)Wang, Wang, \& Han}]{wang2014polarized}
Wang, P., Wang, C., \& Han, J. 2014, Monthly Notices of the Royal Astronomical
  Society, 441, 1943

\bibitem[{{Wang} {et~al.}(2020){Wang}, {Xu}, \& {Chen}}]{2020ApJ...899..109W}
{Wang}, W.-Y., {Xu}, R., \& {Chen}, X. 2020, \apj, 899, 109,
  \dodoi{10.3847/1538-4357/aba268}

\bibitem[{{Wang} {et~al.}(2022){Wang}, {Yang}, {Niu}, {Xu}, \&
  {Zhang}}]{2022ApJ...927..105W}
{Wang}, W.-Y., {Yang}, Y.-P., {Niu}, C.-H., {Xu}, R., \& {Zhang}, B. 2022,
  \apj, 927, 105, \dodoi{10.3847/1538-4357/ac4097}

\bibitem[{{Wharton} {et~al.}(2021){Wharton}, {Bethapudi}, {Marthi}, {Main},
  {Li}, {Gautam}, {Lin}, {Spitler}, \& {Pen}}]{Wharton2021ATel14538}
{Wharton}, R., {Bethapudi}, S., {Marthi}, V., {et~al.} 2021, The Astronomer's
  Telegram, 14538, 1

\bibitem[{Xilouris {et~al.}(1995)Xilouris, Seiradakis, Gil, Sieber, \&
  Wielebinski}]{xilouris1995pulsar}
Xilouris, K., Seiradakis, J., Gil, J., Sieber, W., \& Wielebinski, R. 1995,
  Astronomy and Astrophysics, Vol. 293, p. 153-165 (1995), 293, 153

\bibitem[{Xu {et~al.}(2022)Xu, Niu, Chen, Lee, Zhu, Dong, Zhang, Jiang, Wang,
  Xu, {et~al.}}]{xu2022fast}
Xu, H., Niu, J., Chen, P., {et~al.} 2022, Nature, 609, 685

\bibitem[{{Xu} {et~al.}(1997){Xu}, {Qiao}, \& {Han}}]{1997A&A...323..395X}
{Xu}, R.~X., {Qiao}, G.~J., \& {Han}, J.~L. 1997, \aap, 323, 395

\bibitem[{{Yang} \& {Zhang}(2018)}]{2018ApJ...868...31Y}
{Yang}, Y.-P., \& {Zhang}, B. 2018, \apj, 868, 31,
  \dodoi{10.3847/1538-4357/aae685}

\bibitem[{{Zhang}(2022)}]{2022ApJ...925...53Z}
{Zhang}, B. 2022, \apj, 925, 53, \dodoi{10.3847/1538-4357/ac3979}

\bibitem[{Zhang(2023)}]{ZhangB2023}
Zhang, B. 2023, Reviews of Modern Physics, 95, 035005

\bibitem[{Zhang {et~al.}(2022)Zhang, Wang, Feng, Zhang, Li, Tsai, Niu, Luo,
  Yao, Zhu, {et~al.}}]{zhang2022fast}
Zhang, Y.-K., Wang, P., Feng, Y., {et~al.} 2022, Research in Astronomy and
  Astrophysics, 22, 124002

\bibitem[{Zhou {et~al.}(2022)Zhou, Han, Zhang, Lee, Zhu, Li, Jing, Wang, Zhang,
  Jiang, {et~al.}}]{zhou2022fast}
Zhou, D., Han, J., Zhang, B., {et~al.} 2022, Research in Astronomy and
  Astrophysics, 22, 124001

\end{thebibliography}
\bibliographystyle{aasjournal}



\end{document}